\documentclass[sigconf]{acmart}

\AtBeginDocument{%
  \providecommand\BibTeX{{%
    \normalfont B\kern-0.5em{\scshape i\kern-0.25em b}\kern-0.8em\TeX}}}

\copyrightyear{2024}
\acmYear{2024}
\setcopyright{rightsretained}
\acmConference[CHI '24 Computational UI Workshop]{CHI Conference on Human Factors in Computing Systems}{May 12, 2024}{Honolulu, HI, USA}

\usepackage{multirow}
\usepackage{makecell}

\begin{document}

\title{Unveiling Disparities in Web Task Handling Between Human and Web Agent}

\author{Kihoon Son}
\email{kihoon.son@kaist.ac.kr}
\affiliation{%
  \institution{School of Computing, KAIST}
  \city{Daejeon}
  \country{Republic of Korea}
}
\author{Jinhyeon Kwon}
\email{athexplorer@kaist.ac.kr}
\affiliation{%
  \institution{School of Computing, KAIST}
  \city{Daejeon}
  \country{Republic of Korea}
}
\author{DaEun Choi}
\email{daeun.choi@kaist.ac.kr}
\affiliation{%
  \institution{School of Computing, KAIST}
  \city{Daejeon}
  \country{Republic of Korea}
}
\author{Tae Soo Kim}
\email{taesoo.kim@kaist.ac.kr}
\affiliation{%
  \institution{School of Computing, KAIST}
  \city{Daejeon}
  \country{Republic of Korea}
}
\author{Young-Ho Kim}
\email{yghokim@younghokim.net}
\affiliation{%
  \institution{NAVER AI Lab}
  \city{Seongnam}
  \country{Republic of Korea}
}
\author{Sangdoo Yun}
\email{sangdoo.yun@navercorp.com}
\affiliation{%
  \institution{NAVER AI Lab}
  \city{Seongnam}
  \country{Republic of Korea}
}
\author{Juho Kim}
\email{juhokim@kaist.ac.kr}
\affiliation{%
  \institution{School of Computing, KAIST}
  \city{Daejeon}
  \country{Republic of Korea}
}
\renewcommand{\shortauthors}{Kihoon Son, Jinhyeon Kwon, DaEun Choi, Tae Soo Kim, Young-Ho Kim, Sangdoo Yun, and Juho Kim}

\begin{abstract}
With the advancement of Large-Language Models (LLMs) and Large Vision-Language Models (LVMs), agents have shown significant capabilities in various tasks, such as data analysis, gaming, or code generation. Recently, there has been a surge in research on web agents, capable of performing tasks within the web environment. However, the web poses unforeseeable scenarios, challenging the generalizability of these agents. This study investigates the disparities between human and web agents' performance in web tasks (e.g., information search) by concentrating on planning, action, and reflection aspects during task execution. We conducted a web task study with a think-aloud protocol, revealing distinct cognitive actions and operations on websites employed by humans. Comparative examination of existing agent structures and human behavior with thought processes highlighted differences in knowledge updating and ambiguity handling when performing the task. Humans demonstrated a propensity for exploring and modifying plans based on additional information and investigating reasons for failure. These findings offer insights into designing planning, reflection, and information discovery modules for web agents and designing the capturing method for implicit human knowledge in a web task.\end{abstract}



\maketitle

\section{Introduction}

As Large-Language Models (e.g., GPT-4 ~\cite{openai2023gpt4}) and Large Vision-Language Models (LVM; GPT-4V~\cite{2023GPT4VisionSC}) continue to advance, research on agents has shown that agents are capable of performing diverse tasks to support humans as well as in diverse formats (e.g., single or multi agents framework). For example, LLM-based agents perform tasks such as data analysis~\cite{Hu2024InfiAgentDABenchEA}, gaming~\cite{Zhu2023CALYPSOLA}, and code generation~\cite{Zhu2023CALYPSOLA}. Also, recent agent work leverages a multi-agent framework with diverse personas to support human decision-making processes~\cite{Wu2023AutoGenEN, Park2023ChoiceMatesSU}. These LLM-based agents typically incorporate four fundamental components: Profiling, Memory, Planning, and Action~\cite{Wang2023ASO}. Agents with this architecture operate by designing plans for tasks akin to humans, retrieving necessary information from knowledge memory, and subsequently executing the most appropriate actions based on the current context~\cite{Wu2024OSCopilotTG}.

Recently, with the capability of understanding graphic user interface (GUI), web agents based on LLM and LVM have shown to be capable of performing tasks within the web environment~\cite{Koh2024VisualWebArenaEM}, such as shopping or finding specific information in a social community. To successfully accomplish tasks on the web, these agents require not only a basic understanding of graphic user interface (GUI) but also appropriate planning on how to perform given tasks and accurately execute an action ~\cite{Cheng2024SeeClickHG}. Researchers have shown high interest in how to provide the visual grounding~\cite{Yang2023SetofMarkPU, Zheng2024GPT4VisionIA} to the agent and generate a task plan with sub-plans~\cite{Niu2024ScreenAgentAV} and how to design an agent structure ~\cite{Niu2024ScreenAgentAV}.

However, the web environment presents many unforeseeable challenges (e.g., encountering irrelevant search results or unconventional UI and interaction design), leading to a decrease in the generalizability of agents in web tasks~\cite{Niu2024ScreenAgentAV}. When faced with unexpected scenarios, humans contemplate reasons and seek alternative solutions while planning for the next action. However, current web agents lack modules capable of performing such functions.
Moreover, while humans can perform given tasks on unfamiliar websites through basic UI understanding and trial and error, web agents still lack this ability. To address this limitation, ScreenAgent~\cite{Niu2024ScreenAgentAV} has introduced a module capable of reflecting on its own actions, and FRIDAY~\cite{Wu2024OSCopilotTG} has proposed module designs considering the biological nature of the human brain to make an agent behave like a human in performing the task. However, the low performance of these agents might imply the existence of implicit knowledge or behavior patterns unique to humans on the web task to be unpacked.

Then, what are the differences between humans and web agents in performing web tasks? What distinguishes agents from humans in terms of planning, action, and reflection when executing web tasks? Furthermore, what implications can be derived from these differences for agent design? To answer these questions, we conducted a think-aloud study where four participants conducted two different web tasks typically executed by current web agents, such as shopping and finding information about a post. Through a qualitative coding analysis on the think-aloud transcripts, we identified the human's cognitive actions (Table~\ref{tab:cognitive}) and operations on the web (Table ~\ref{tab:operation}). The tables illustrate the types of planning humans make, the decisions they make, and how they perform rollback in plans when something goes wrong in web tasks. 

Furthermore, by comparing participants' task execution processes with ScreenAgent~\cite{Niu2024ScreenAgentAV}'s prompting structure and its modules, we identified that humans update their knowledge space (task- and task website-related knowledge) by exploring and discovering new information. Our results also reveal that humans exhibit a pattern of identifying and clarifying ambiguous aspects of tasks through an additional information search beyond simply performing tasks according to constructed plans. Unlike agents, humans investigate reasons for failure or discrepancies between their knowledge and the information they are viewing, and based on this exploration, they modify their plans. Based on these findings, we discuss design implications for additional modules needed in agents and how to capture implicit knowledge and behavior patterns unique to humans during task execution processes. Our discussion also includes the methods for capturing humans' tacit knowledge within the web environment to transfer this knowledge to the agent.



\section{Main Study}
To investigate how humans execute web tasks in depth, we conducted a think-aloud study with four participants on the two different web tasks.

\subsection{Study Design}
\subsubsection{Task design}
The tasks of study are a shopping task on the Amazon website~\footnote{https://www.amazon.com/} and an information search task on Reddit~\footnote{https://www.reddit.com/}. We selected the tasks based on the VisualWebArena paper~\cite{Koh2024VisualWebArenaEM} since these are the tasks commonly used in web agent research. The two tasks are:
\begin{itemize}
    \item Find the cheapest Lavazza Gran Aroma coffee capsule on Amazon.
    \item Find the title of the\_Champion’s \textit{Hot post} at r/midjourney ten months ago.
\end{itemize}

\subsubsection{Participant}
We recruited a total of four participants. Two people are familiar with the task and the given websites, and the other two people aren't. This recruitment would allow us to observe how people unfamiliar with the website perceive, understand, and respond to the given task.

\subsubsection{Procedure}
The study started with a brief introduction by the authors, followed by the execution of the two tasks by the participants. We instructed them to think aloud in terms of their planning, action, and reflection. Specifically, we asked them to explain the rationale or intent behind their task planning and action. Then, the participants conducted the two tasks in random order until they completed or quit the given task. After conducting the two tasks, the study finished.

\subsection{Analysis}
To analyze the think-aloud transcript data, we conducted an open coding ~\cite{corbin1990grounded, khandkar2009open} on the think-aloud transcript data. 

First, one author organized the transcripts while monitoring recorded study sessions. In cases where verbal explanations lacked context (such as instances involving pronouns like "this" and "that" in the think-aloud process), the author checked the specific situation from the recorded video. Through this process, we obtained 156 raw data from transcripts in all study sessions.

Second, the two authors jointly conducted all stages of qualitative coding. Without further interpretation, the authors initially classified the raw data into similar codes first and defined names for the codes. Next, the authors reviewed the defined codes together, merging similar codes and splitting codes if different types of raw data existed within the same code. The entire process continued until a consensus was reached between the two authors. 

Through this process, the 156 transcript raw data were classified into 39 codes. The two authors then defined 14 themes by grouping similar codes again. Similarly, consensus was sought between the two authors in the process of grouping and separating themes. Finally, the grouped themes were divided into cognitive aspects of action (Table \ref{tab:cognitive}) and operational (Table \ref{tab:operation}) aspects performed on the actual website when humans perform web tasks.

\begin{figure*}[!ht]
  \centering
  \includegraphics[width=0.9\linewidth]{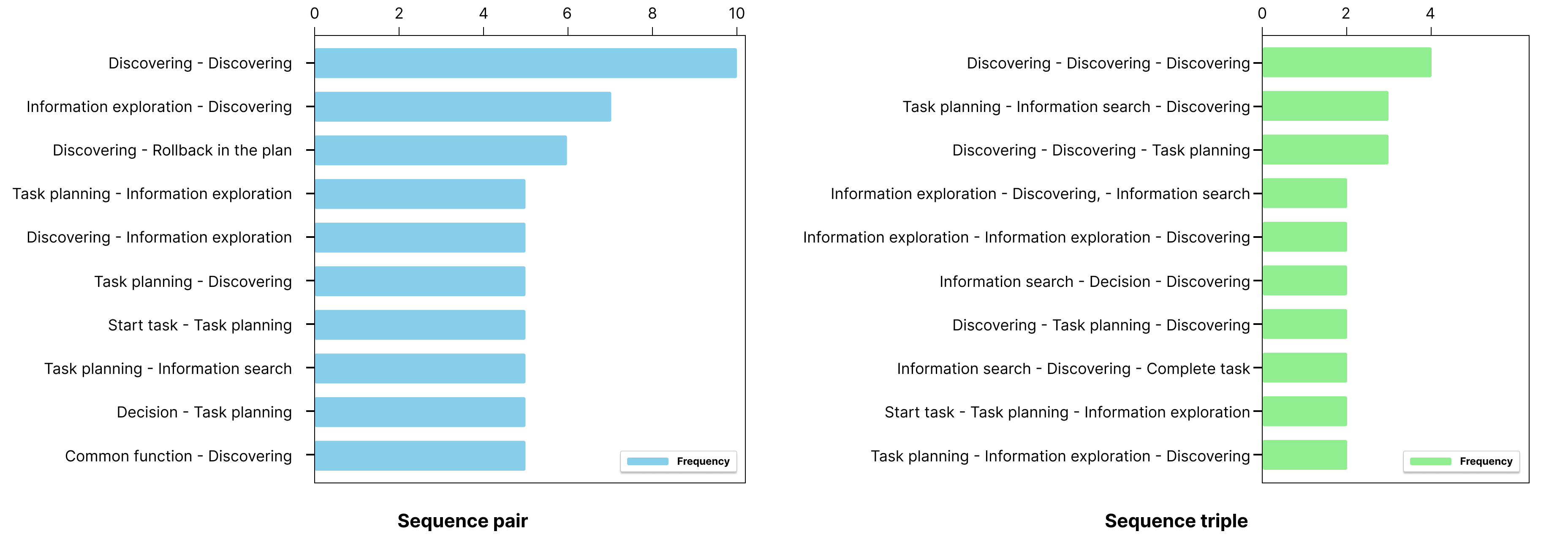}
  \caption{Analysis results of participant sequence data pairs and triples based on the themes of the qualitative coding.}
  \label{fig:sequence}
\end{figure*}

Finally, by assigning the themes determined through qualitative coding back to the original raw data, we created task action sequences of cognitive actions and operations performed by participants during web tasks. In these sequences, each participant's think-aloud data was assigned the themes in Table \ref{tab:cognitive} and \ref{tab:operation}. From a total of 8 sequences (4 participants and 2 tasks), we further analyzed the most prevalent patterns by calculating sequence pairs and triples (Figure \ref{fig:sequence}).
\section{Results}
This section illustrates the results of the analysis of the think-aloud study. Table \ref{tab:cognitive} and Table \ref{tab:operation} show the human's cognitive actions and operation on the web observed in the study. Based on these results, we investigated the core difference between the recent agents.

\begin{table*}
\caption{Qualitative coding results regarding human cognitive actions in performing web tasks.}
\label{tab:cognitive}
\def\arraystretch{1.4}%
\resizebox{\linewidth}{!}{%
\begin{tabular}{lllcc}
\cmidrule[1.2pt]{1-5}
\multicolumn{1}{l}{} 
 & Theme & Code & Count & Ratio \\
\cmidrule[1.2pt]{1-5}
\multirow{23}{*}{\textbf{Cognitive action}} &
  \multirow{4}{*}{\textbf{Task planning}} &
  - Establishing a non-specific initial task plan & \multirow{4}{*}{\textbf{24}} & \multirow{4}{*}{\textbf{15.38\%}} \\
     & & - Establishing an additional auxiliary plan to know unknown information & & \\
     & & - Establishing a verification plan for discerning the truth of the found information or answer & & \\
     & & - Establishing an alternative plan & & \\
  \cmidrule[0.5pt]{2-5}
 &
  
  \multirow{2}{*}{\textbf{Task condition checking}} & - Recalling the task conditions  & \multirow{2}{*}{\textbf{6}} & \multirow{2}{*}{\textbf{3.58\%}}  \\
     & & - Identifying ambiguous parts in the task conditions & & \\
  \cmidrule[0.5pt]{2-5}
 &
  
  \multirow{3}{*}{\textbf{Information processing}} & - Collecting information helpful for performing the task & \multirow{3}{*}{\textbf{3}} & \multirow{3}{*}{\textbf{1.92\%}}\\
     & & - Comparing derived answers and newly discovered information & & \\
     & & - Comparing prior knowledge and currently observed information & & \\
  \cmidrule[0.5pt]{2-5}
 &
 
  \multirow{3}{*}{\textbf{Discovering}} & - Discovering task-related knowledge from currently observed information & \multirow{3}{*}{\textbf{41}} & \multirow{3}{*}{\textbf{26.28\%}}\\
 & & - Discovering task website-related knowledge (e.g., function) through UI understanding, intuition, or trial. & & \\
 & & - Discovering that the current plan is flawed based on the currently observed information & & \\
  \cmidrule[0.5pt]{2-5}
 &
  
  \multirow{4}{*}{\textbf{Decision}} & - Deciding on the information to search & \multirow{4}{*}{\textbf{14}} & \multirow{4}{*}{\textbf{8.97\%}}\\
 & & - Deciding on answer candidates & & \\
 & & - Deciding on whether to maintain the current plan & & \\
 & & - Deciding on defining the ambiguous parts in task conditions & & \\
  \cmidrule[0.5pt]{2-5}
 &

  \multirow{4}{*}{\textbf{Complete task}} & - Completing the task through comparison of currently observed information with the derived answers & \multirow{4}{*}{\textbf{7}}& \multirow{4}{*}{\textbf{4.49\%}}\\
 & & - Completing the task by confirming satisfaction of task conditions & & \\
 & & - Completing the task based on prior knowledge of the task website & & \\
 & & - Completing the task by clarifying the vague condition & & \\
  \cmidrule[0.5pt]{2-5}
 &
 
  \multirow{3}{*}{\textbf{Reflection}} & - Contemplating the reasons for the current plan's failure &\multirow{3}{*}{\textbf{6}} &\multirow{3}{*}{\textbf{3.85\%}} \\
 & & - Contemplating the differences between prior knowledge and newly found information & & \\
 & & - Contemplating whether there has been an information gain from the information currently being viewed & & \\
  \cmidrule[1.2pt]{1-5}
\end{tabular}
}
\end{table*}

\begin{table*}
\caption{Qualitative coding results regarding human operations on a website in performing web tasks.}
\label{tab:operation}
\def\arraystretch{1.4}%
\resizebox{\linewidth}{!}{%
\begin{tabular}{p{0.2\columnwidth}p{0.4\columnwidth}lcc}
\cmidrule[1.2pt]{1-5}
\multicolumn{1}{l}{} & Theme & Code & Count & Ratio\\
\cmidrule[1.2pt]{1-5}
\multirow{16}{*}{\textbf{Operation}} & \multirow{3}{*}{\textbf{Information search}} & - Searching the given conditions & \multirow{3}{*}{\textbf{12}} & \multirow{3}{*}{\textbf{7.69\%}} \\
 &  & - Search for specific information directly related to answers & &  \\
 &  & - Search for clarification of unknown part in Google & &  \\
 \cmidrule[0.5pt]{2-5}
 & \multirow{4}{*}{\textbf{Information exploration}} & - Exploring specific information by considering the task conditions or unclear parts & \multirow{4}{*}{\textbf{16}} & \multirow{4}{*}{\textbf{10.26\%}} \\
 &  & - Exploring information to verify its irrelevance, despite initial judgment of it being incorrect  & & \\
 &  & - Exploring the website functionalities at the beginning of the task  & & \\
 &  & - Exploring previously opened tabs that might contain useful information  & & \\
 \cmidrule[0.5pt]{2-5}
 & \multirow{4}{*}{\textbf{Roll-back in the plan}} & - Roll-backing from auxiliary plan to auxiliary plan  &\multirow{4}{*}{\textbf{9}} & \multirow{4}{*}{\textbf{5.77\%}}\\
 &  & - Roll-backing from auxiliary plan to main plan  & & \\
 &  & - Roll-backing to start point  & & \\
 &  & - Roll-backing from main plan to main plan  & & \\
 \cmidrule[0.5pt]{2-5}
 & \multirow{2}{*}{\textbf{Start task}} & - Starting task with prior knowledge like prefix of website address  &\multirow{2}{*}{\textbf{7}} & \multirow{2}{*}{\textbf{4.49\%}}\\
 &  & - Starting the task by navigating to the task website through a Google search  & &\\
 \cmidrule[0.5pt]{2-5}
 & \textbf{Trial and error} & - Trying despite uncertainty about the success of the action  & \multirow{1}{*}{\textbf{2}}& \multirow{1}{*}{\textbf{1.28\%}} \\
 \cmidrule[0.5pt]{2-5}
 & \textbf{General function} & - Click on the \textit{next step}, \textit{filtering}, or \textit{sorting}  & \multirow{1}{*}{\textbf{7}}& \multirow{1}{*}{\textbf{4.49\%}} \\
 \cmidrule[0.5pt]{2-5}
 & \textbf{Task specific function} & - Specific task website functions like \textit{Add to cart}  & \multirow{1}{*}{\textbf{2}}& \multirow{1}{*}{\textbf{1.28\%}}\\
 \cmidrule[1.2pt]{1-5}
\end{tabular}
}
\end{table*}

\subsection{Cognitive Actions and Website Operations}
Qualitative coding result (\autoref{tab:cognitive}) reveals that in performing web tasks, humans engage in various cognitive actions beyond task planning and reflection. Particularly, during the task process, humans frequently discover task-related knowledge from the information they are currently observing or uncover functionalities of the website previously unknown to them through UI understanding (e.g., catching a website's functionality through UI design) and intuition. Task-related knowledge refers to information unknown within task conditions (e.g., the fact that Lavazza capsules come in various types), while task website functionalities denote commands or features used within the website (e.g., how to find a specific username on Reddit using \textit{u/}). Participants familiar with the task utilized such information as prior knowledge, whereas those unfamiliar with the task progressed through exploration to clarify these unclear aspects while performing the task.

From the operational aspect of tasks performed on websites (\autoref{tab:operation}), it is evident that humans engage in search and exploration operations more frequently than others when encountering unknown information. Despite participants' high familiarity with website functions, operations such as search and exploration were commonly employed to clarify new information within the given conditions (e.g., types of Lavazza coffee capsules). For instance, in a task requiring participants to find coffee capsules, even those familiar with the website resorted to searching for related information to understand the difference between Lavazza's pot and coffee capsules, thus enabling them to make a clear selection based on Amazon search results. Additionally, participants tended to establish additional auxiliary plans for exploring the unknown aspects rather than entirely redesigning their task plans. They also demonstrated a propensity to formulate plans for verifying newly acquired information. Moreover, during task execution, when the direction of exploration went away or when it was unclear which operation to perform, individuals exhibited a roll-back operation within the established plan, allowing them to revert to a specific point.

\subsection{Comparing Web Agent and Human}

\subsubsection{Human's Discovering Action and Knowledge Update Process in Web Task}
Our study results show that the participants acquired knowledge relevant to tasks (e.g., the difference between the types of coffee capsules) and task websites (e.g., website function or search results pattern) through their own \textit{UI understanding}, \textit{trial and error}, or \textit{intuition} while performing tasks. Furthermore, they frequently validate whether their current plan is correct or not through these discoveries. Through this process of discovery, participants perform rollback within their plan or additional information exploration operations. On the cognitive action aspect, participants made decisions related to whether to perform reflection based on the information discovered, conduct information searches or exploration, determine answers, or maintain the current plan.

To update the discovered knowledge during the task, the current structure of web agents requires a framework that can distinguish and update knowledge about the given task itself and knowledge about the task website based on the information discovered. In the process of discovering information, web agents might need a structure that can capture and reflect the above information in the next decision based on task context or general website knowledge (Website structure or UI) gained from past experiences.

\subsubsection{Information Exploration with Trial and Error}
Current web agents mainly execute actions that correspond to sub-plans once a plan is formulated~\cite{Niu2024ScreenAgentAV}. From observing participants unfamiliar with the task, we noted that they sought to clarify vague parts of the task through exploration. An interesting point is that participants not only explore potentially useful information but also the functionality of the task website and sometimes inappropriate information. Ultimately, after the exploration, they make decisions about which information is useful and which is not. This allows the participants to expand their knowledge space related to the task- or task website-related knowledge.

\subsubsection{Differences of Reflection on What}
ScreenAgent ~\cite{Niu2024ScreenAgentAV} executes a specific action for a given sub-plan and performs reflection by observing changes on the screen before and after execution. Through this process, it either 1) repeats the action, 2) revises the plan, or 3) proceeds to the next step.

In contrast to this structure, our study results indicate that humans engage in a more complex reflection process. Humans reflect not only on why a plan failed but also on how newly acquired information differs from their existing knowledge and whether the information gained adds new value to their knowledge space beyond simply understanding why a plan with an action failed.

Based on this reflection, humans revise plans, perform rollbacks within plans, or determine the direction of information exploration within the knowledge space. Unlike current web agents, humans extend their reflection beyond evaluating whether actions performed for sub-plans were executed correctly. They delve into reasons for failure and possibilities for remediation and identify additional information that may require exploration.
\section{Discussion}
Based on the study's results, this section explains how web agents would be designed and discusses the limitations of think-aloud studies in capturing implicit human behavior and knowledge.

\subsection{Design Implications for Web Agent}
Web agents must possess not only knowledge about the given task but also the necessary knowledge within the task website. In the study, participants demonstrated updating both forms of knowledge, unlike web agents. Beyond simply providing task instructions as prompts, there is a need for modules capable of understanding knowledge about the tasks themselves, comparing them with new information discovered during tasks, and updating accordingly. Additionally, to respond to various forms and designs of website environments appropriately, web agents would require a combination of general web knowledge applicable across websites and specialized web knowledge tailored to specific websites or tasks.

In addition to incorporating modules related to knowledge, it may be possible to design modules within the agent for human processes such as reflection, planning, roll-back, and decision-making. The relationships between these processes and how they interact during task execution may require further exploration, depending on the complexity of the tasks. However, initially, it would be essential to delineate the processes present in humans and design them as modules within the agent. Research into how these newly designed modules could enhance the current performance of the agent would be necessary.

\subsection{Capturing Human's Tacit Knowledge}
In our study, we observed that people vary in the depth of their explanations level, and it may not comprehensively capture all processes with the think-aloud protocol ~\cite{krosnick2021think}. Although the study tasks were simpler than open-ended tasks without predefined answers, such as a literature survey or design ideation, we observed that participants missed certain aspects when explaining their thoughts and that the level of details of explanation varied between individuals. Particularly, it was challenging for participants to provide clear explanations for how their current actions or explanations related to their previous explanations, behaviors, or their prior knowledge. From the behaviors and thought processes unique to humans, such as the process of acquiring new information or performing tasks based on it, there is a need to improve the think-aloud protocol to clearly capture aspects applicable to agents. 

Therefore, in this background, it is important to capture the human's implicit behavior pattern or tacit knowledge within a certain task process to reuse the task-related knowledge~\cite{son2024demystifying}. This might also reveal a more detailed implicit process, like how humans conduct the trial-and-error processes or revise what parts of their plan are in a certain situation. Furthermore, if we can create datasets containing even implicit patterns or tacit knowledge and feed them to train and fine-tune agents, they would be able to perform tasks more like humans and offer a wider range of support to people. Through the dataset, to provide general or personalized support within the task, we can also define the ground rules of the task knowledge that various people utilized during the task, as well as the personalized rules or knowledge.
\section{Conclusion}
In conclusion, this study sheds light on the disparities observed between human and web agents' performance in web tasks. Through a think-aloud study of web tasks, distinct cognitive actions and operations on websites utilized by humans were revealed. Comparative analysis with existing agent structures underscored differences in knowledge updating and ambiguity handling during task execution. Humans demonstrated a propensity for adaptive planning based on additional information and thorough investigation into reasons for failure. These findings provide design insights into web agent architectures and the development of methods for capturing implicit human knowledge within web tasks. By capturing and transforming implicit behavior patterns with tacit knowledge utilized by individuals during task execution into forms applicable to web agents, we envision a future where web agents can provide more diverse support for humans.
\begin{acks}
This work was supported by NAVER-KAIST Hypercreative AI Center. We thank all of our participants for engaging positively in our studies. We also thank all of the members of KIXLAB for their helpful discussions and constructive feedback.    
\end{acks}

\bibliographystyle{ACM-Reference-Format}
\bibliography{references}


\begin{thebibliography}{17}


\ifx \showCODEN    \undefined \def \showCODEN     #1{\unskip}     \fi
\ifx \showDOI      \undefined \def \showDOI       #1{#1}\fi
\ifx \showISBNx    \undefined \def \showISBNx     #1{\unskip}     \fi
\ifx \showISBNxiii \undefined \def \showISBNxiii  #1{\unskip}     \fi
\ifx \showISSN     \undefined \def \showISSN      #1{\unskip}     \fi
\ifx \showLCCN     \undefined \def \showLCCN      #1{\unskip}     \fi
\ifx \shownote     \undefined \def \shownote      #1{#1}          \fi
\ifx \showarticletitle \undefined \def \showarticletitle #1{#1}   \fi
\ifx \showURL      \undefined \def \showURL       {\relax}        \fi
\providecommand\bibfield[2]{#2}
\providecommand\bibinfo[2]{#2}
\providecommand\natexlab[1]{#1}
\providecommand\showeprint[2][]{arXiv:#2}

\bibitem[202(2023)]%
        {2023GPT4VisionSC}
 \bibinfo{year}{2023}\natexlab{}.
\newblock \showarticletitle{GPT-4V(ision) System Card}.
\newblock
\urldef\tempurl%
\url{https://api.semanticscholar.org/CorpusID:263218031}
\showURL{%
\tempurl}


\bibitem[Cheng et~al\mbox{.}(2024)]%
        {Cheng2024SeeClickHG}
\bibfield{author}{\bibinfo{person}{Kanzhi Cheng}, \bibinfo{person}{Qiushi Sun}, \bibinfo{person}{Yougang Chu}, \bibinfo{person}{Fangzhi Xu}, \bibinfo{person}{Yantao Li}, \bibinfo{person}{Jianbing Zhang}, {and} \bibinfo{person}{Zhiyong Wu}.} \bibinfo{year}{2024}\natexlab{}.
\newblock \showarticletitle{SeeClick: Harnessing GUI Grounding for Advanced Visual GUI Agents}.
\newblock \bibinfo{journal}{\emph{ArXiv}}  \bibinfo{volume}{abs/2401.10935} (\bibinfo{year}{2024}).
\newblock
\urldef\tempurl%
\url{https://api.semanticscholar.org/CorpusID:267069082}
\showURL{%
\tempurl}


\bibitem[Corbin and Strauss(1990)]%
        {corbin1990grounded}
\bibfield{author}{\bibinfo{person}{Juliet Corbin} {and} \bibinfo{person}{Anselm Strauss}.} \bibinfo{year}{1990}\natexlab{}.
\newblock \showarticletitle{Grounded Theory Research: Procedures, Canons and Evaluative Criteria}.
\newblock \bibinfo{journal}{\emph{Zeitschrift für Soziologie}} \bibinfo{volume}{19}, \bibinfo{number}{6} (\bibinfo{date}{Dec.} \bibinfo{year}{1990}), \bibinfo{pages}{418–427}.
\newblock
\showISSN{0340-1804}
\urldef\tempurl%
\url{https://doi.org/10.1515/zfsoz-1990-0602}
\showDOI{\tempurl}


\bibitem[Hu et~al\mbox{.}(2024)]%
        {Hu2024InfiAgentDABenchEA}
\bibfield{author}{\bibinfo{person}{Xueyu Hu}, \bibinfo{person}{Ziyu Zhao}, \bibinfo{person}{Shuang Wei}, \bibinfo{person}{Ziwei Chai}, \bibinfo{person}{Guoyin Wang}, \bibinfo{person}{Xuwu Wang}, \bibinfo{person}{Jing Su}, \bibinfo{person}{Jingjing Xu}, \bibinfo{person}{Ming Zhu}, \bibinfo{person}{Yao Cheng}, \bibinfo{person}{Jianbo Yuan}, \bibinfo{person}{Kun Kuang}, \bibinfo{person}{Yang Yang}, \bibinfo{person}{Hongxia Yang}, {and} \bibinfo{person}{Fei Wu}.} \bibinfo{year}{2024}\natexlab{}.
\newblock \showarticletitle{InfiAgent-DABench: Evaluating Agents on Data Analysis Tasks}.
\newblock \bibinfo{journal}{\emph{ArXiv}}  \bibinfo{volume}{abs/2401.05507} (\bibinfo{year}{2024}).
\newblock
\urldef\tempurl%
\url{https://api.semanticscholar.org/CorpusID:266933185}
\showURL{%
\tempurl}


\bibitem[Khandkar(2009)]%
        {khandkar2009open}
\bibfield{author}{\bibinfo{person}{Shahedul~Huq Khandkar}.} \bibinfo{year}{2009}\natexlab{}.
\newblock \showarticletitle{Open coding}.
\newblock \bibinfo{journal}{\emph{University of Calgary}}  \bibinfo{volume}{23} (\bibinfo{year}{2009}), \bibinfo{pages}{2009}.
\newblock
\urldef\tempurl%
\url{http://pages.cpsc.ucalgary.ca/~saul/wiki/uploads/CPSC681/opencoding.pdf}
\showURL{%
\tempurl}


\bibitem[Koh et~al\mbox{.}(2024)]%
        {Koh2024VisualWebArenaEM}
\bibfield{author}{\bibinfo{person}{Jing~Yu Koh}, \bibinfo{person}{Robert Lo}, \bibinfo{person}{Lawrence Jang}, \bibinfo{person}{Vikram Duvvur}, \bibinfo{person}{Ming~Chong Lim}, \bibinfo{person}{Po-Yu Huang}, \bibinfo{person}{Graham Neubig}, \bibinfo{person}{Shuyan Zhou}, \bibinfo{person}{Ruslan Salakhutdinov}, {and} \bibinfo{person}{Daniel Fried}.} \bibinfo{year}{2024}\natexlab{}.
\newblock \showarticletitle{VisualWebArena: Evaluating Multimodal Agents on Realistic Visual Web Tasks}.
\newblock \bibinfo{journal}{\emph{ArXiv}}  \bibinfo{volume}{abs/2401.13649} (\bibinfo{year}{2024}).
\newblock
\urldef\tempurl%
\url{https://api.semanticscholar.org/CorpusID:267199749}
\showURL{%
\tempurl}


\bibitem[Krosnick et~al\mbox{.}(2021)]%
        {krosnick2021think}
\bibfield{author}{\bibinfo{person}{Rebecca Krosnick}, \bibinfo{person}{Fraser Anderson}, \bibinfo{person}{Justin Matejka}, \bibinfo{person}{Steve Oney}, \bibinfo{person}{Walter S.~Lasecki}, \bibinfo{person}{Tovi Grossman}, {and} \bibinfo{person}{George Fitzmaurice}.} \bibinfo{year}{2021}\natexlab{}.
\newblock \showarticletitle{Think-Aloud Computing: Supporting Rich and Low-Effort Knowledge Capture}. In \bibinfo{booktitle}{\emph{Proceedings of the 2021 CHI Conference on Human Factors in Computing Systems}} (<conf-loc>, <city>Yokohama</city>, <country>Japan</country>, </conf-loc>) \emph{(\bibinfo{series}{CHI '21})}. \bibinfo{publisher}{Association for Computing Machinery}, \bibinfo{address}{New York, NY, USA}, Article \bibinfo{articleno}{199}, \bibinfo{numpages}{13}~pages.
\newblock
\showISBNx{9781450380966}
\urldef\tempurl%
\url{https://doi.org/10.1145/3411764.3445066}
\showDOI{\tempurl}


\bibitem[Niu et~al\mbox{.}(2024)]%
        {Niu2024ScreenAgentAV}
\bibfield{author}{\bibinfo{person}{Runliang Niu}, \bibinfo{person}{Jindong Li}, \bibinfo{person}{Shiqi Wang}, \bibinfo{person}{Yali Fu}, \bibinfo{person}{Xiyu Hu}, \bibinfo{person}{Xueyuan Leng}, \bibinfo{person}{He Kong}, \bibinfo{person}{Yi Chang}, {and} \bibinfo{person}{Qi Wang}.} \bibinfo{year}{2024}\natexlab{}.
\newblock \showarticletitle{ScreenAgent: A Vision Language Model-driven Computer Control Agent}.
\newblock \bibinfo{journal}{\emph{ArXiv}}  \bibinfo{volume}{abs/2402.07945} (\bibinfo{year}{2024}).
\newblock
\urldef\tempurl%
\url{https://api.semanticscholar.org/CorpusID:267636776}
\showURL{%
\tempurl}


\bibitem[OpenAI(2023)]%
        {openai2023gpt4}
\bibfield{author}{\bibinfo{person}{OpenAI}.} \bibinfo{year}{2023}\natexlab{}.
\newblock \bibinfo{title}{GPT-4 Technical Report}.
\newblock
\newblock
\showeprint[arxiv]{2303.08774}~[cs.CL]


\bibitem[Park et~al\mbox{.}(2023)]%
        {Park2023ChoiceMatesSU}
\bibfield{author}{\bibinfo{person}{Jeongeon Park}, \bibinfo{person}{Bryan Min}, \bibinfo{person}{Xiaojuan Ma}, {and} \bibinfo{person}{Juho Kim}.} \bibinfo{year}{2023}\natexlab{}.
\newblock \showarticletitle{ChoiceMates: Supporting Unfamiliar Online Decision-Making with Multi-Agent Conversational Interactions}.
\newblock \bibinfo{journal}{\emph{ArXiv}}  \bibinfo{volume}{abs/2310.01331} (\bibinfo{year}{2023}).
\newblock
\urldef\tempurl%
\url{https://api.semanticscholar.org/CorpusID:263605868}
\showURL{%
\tempurl}


\bibitem[Son et~al\mbox{.}(2024)]%
        {son2024demystifying}
\bibfield{author}{\bibinfo{person}{Kihoon Son}, \bibinfo{person}{DaEun Choi}, \bibinfo{person}{Tae~Soo Kim}, {and} \bibinfo{person}{Juho Kim}.} \bibinfo{year}{2024}\natexlab{}.
\newblock \bibinfo{title}{Demystifying Tacit Knowledge in Graphic Design: Characteristics, Instances, Approaches, and Guidelines}.
\newblock
\newblock
\showeprint[arxiv]{2403.06252}~[cs.HC]


\bibitem[Wang et~al\mbox{.}(2023)]%
        {Wang2023ASO}
\bibfield{author}{\bibinfo{person}{Lei Wang}, \bibinfo{person}{Chengbang Ma}, \bibinfo{person}{Xueyang Feng}, \bibinfo{person}{Zeyu Zhang}, \bibinfo{person}{Hao ran Yang}, \bibinfo{person}{Jingsen Zhang}, \bibinfo{person}{Zhi-Yang Chen}, \bibinfo{person}{Jiakai Tang}, \bibinfo{person}{Xu Chen}, \bibinfo{person}{Yankai Lin}, \bibinfo{person}{Wayne~Xin Zhao}, \bibinfo{person}{Zhewei Wei}, {and} \bibinfo{person}{Ji rong Wen}.} \bibinfo{year}{2023}\natexlab{}.
\newblock \showarticletitle{A Survey on Large Language Model based Autonomous Agents}.
\newblock \bibinfo{journal}{\emph{ArXiv}}  \bibinfo{volume}{abs/2308.11432} (\bibinfo{year}{2023}).
\newblock
\urldef\tempurl%
\url{https://api.semanticscholar.org/CorpusID:261064713}
\showURL{%
\tempurl}


\bibitem[Wu et~al\mbox{.}(2023)]%
        {Wu2023AutoGenEN}
\bibfield{author}{\bibinfo{person}{Qingyun Wu}, \bibinfo{person}{Gagan Bansal}, \bibinfo{person}{Jieyu Zhang}, \bibinfo{person}{Yiran Wu}, \bibinfo{person}{Shaokun Zhang}, \bibinfo{person}{Erkang Zhu}, \bibinfo{person}{Beibin Li}, \bibinfo{person}{Li Jiang}, \bibinfo{person}{Xiaoyun Zhang}, {and} \bibinfo{person}{Chi Wang}.} \bibinfo{year}{2023}\natexlab{}.
\newblock \showarticletitle{AutoGen: Enabling Next-Gen LLM Applications via Multi-Agent Conversation Framework}.
\newblock \bibinfo{journal}{\emph{ArXiv}}  \bibinfo{volume}{abs/2308.08155} (\bibinfo{year}{2023}).
\newblock
\urldef\tempurl%
\url{https://api.semanticscholar.org/CorpusID:260925901}
\showURL{%
\tempurl}


\bibitem[Wu et~al\mbox{.}(2024)]%
        {Wu2024OSCopilotTG}
\bibfield{author}{\bibinfo{person}{Zhiyong Wu}, \bibinfo{person}{Chengcheng Han}, \bibinfo{person}{Zichen Ding}, \bibinfo{person}{Zhenmin Weng}, \bibinfo{person}{Zhoumianze Liu}, \bibinfo{person}{Shunyu Yao}, \bibinfo{person}{Tao Yu}, {and} \bibinfo{person}{Lingpeng Kong}.} \bibinfo{year}{2024}\natexlab{}.
\newblock \showarticletitle{OS-Copilot: Towards Generalist Computer Agents with Self-Improvement}.
\newblock \bibinfo{journal}{\emph{ArXiv}}  \bibinfo{volume}{abs/2402.07456} (\bibinfo{year}{2024}).
\newblock
\urldef\tempurl%
\url{https://api.semanticscholar.org/CorpusID:267626905}
\showURL{%
\tempurl}


\bibitem[Yang et~al\mbox{.}(2023)]%
        {Yang2023SetofMarkPU}
\bibfield{author}{\bibinfo{person}{Jianwei Yang}, \bibinfo{person}{Hao Zhang}, \bibinfo{person}{Feng Li}, \bibinfo{person}{Xueyan Zou}, \bibinfo{person}{Chun yue Li}, {and} \bibinfo{person}{Jianfeng Gao}.} \bibinfo{year}{2023}\natexlab{}.
\newblock \showarticletitle{Set-of-Mark Prompting Unleashes Extraordinary Visual Grounding in GPT-4V}.
\newblock
\urldef\tempurl%
\url{https://api.semanticscholar.org/CorpusID:264172201}
\showURL{%
\tempurl}


\bibitem[Zheng et~al\mbox{.}(2024)]%
        {Zheng2024GPT4VisionIA}
\bibfield{author}{\bibinfo{person}{Boyuan Zheng}, \bibinfo{person}{Boyu Gou}, \bibinfo{person}{Jihyung Kil}, \bibinfo{person}{Huan Sun}, {and} \bibinfo{person}{Yu Su}.} \bibinfo{year}{2024}\natexlab{}.
\newblock \showarticletitle{GPT-4V(ision) is a Generalist Web Agent, if Grounded}.
\newblock \bibinfo{journal}{\emph{ArXiv}}  \bibinfo{volume}{abs/2401.01614} (\bibinfo{year}{2024}).
\newblock
\urldef\tempurl%
\url{https://api.semanticscholar.org/CorpusID:266741821}
\showURL{%
\tempurl}


\bibitem[Zhu et~al\mbox{.}(2023)]%
        {Zhu2023CALYPSOLA}
\bibfield{author}{\bibinfo{person}{Andrew Zhu}, \bibinfo{person}{Lara~J. Martin}, \bibinfo{person}{Andrew Head}, {and} \bibinfo{person}{Chris Callison-Burch}.} \bibinfo{year}{2023}\natexlab{}.
\newblock \showarticletitle{CALYPSO: LLMs as Dungeon Masters' Assistants}.
\newblock \bibinfo{journal}{\emph{ArXiv}}  \bibinfo{volume}{abs/2308.07540} (\bibinfo{year}{2023}).
\newblock
\urldef\tempurl%
\url{https://api.semanticscholar.org/CorpusID:260900406}
\showURL{%
\tempurl}


\end{thebibliography}

\end{document}